\documentstyle[prb,epsf,twocolumn,aps]{revtex} 

\begin{document}
\draft
\preprint{}

\twocolumn[\hsize\textwidth\columnwidth\hsize\csname
@twocolumnfalse\endcsname

\title
{
Charge dynamics and ``ferromagnetism'' of 
A$_{1-x}$La$_{x}$B$_{6}$ (A=Ca and Sr)
}

\author{K.\ Taniguchi$^{1}$, T.\ Katsufuji$^{1,2}$, 
F.\ Sakai$^{3}$, H.\ Ueda$^{1}$, K.\ Kitazawa$^{1,2}$, and 
H.\ Takagi$^{1}$}
\address{
$^{1}$ Department of Advanced Materials Science, University of 
Tokyo, Tokyo 113-8656, Japan.
}
\address{
$^{2}$ Solution Oriented Research for Science and Technology 
(SORST), Japan Science and Technology Corporation, Kawaguchi 
332-0012, Japan.
}
\address{
$^{3}$ Institute for Solid State Physics, University of Tokyo, 
Kashiwa 277-8581, Japan.
}
\maketitle
\begin{abstract}
Ferromagnetism has been reported recently in La-doped 
alkaline-earth hexaborides, A$_{1-x}$La$_{x}$B$_{6}$ (A=Ca, Sr, 
and Ba). We have performed the reflectivity, Hall resistivity, and 
magnetization measurements of A$_{1-x}$La$_{x}$B$_{6}$.  The 
results indicate that A$_{1-x}$La$_{x}$B$_{6}$ can be regarded as 
a simple doped semimetal, with no signature of an excitonic state as 
suggested by several theories. It is also found that the surface of 
as-grown samples ($\sim$ 10 $\mu$m in thickness) has a different 
electronic structure from a bulk one, and a fairly large number of 
paramagnetic moments are confined in this region. After eliminating 
these paramagnetic moments at the surface, we could not find any 
evidence of an intrinsic ferromagnetic moment in our samples, 
implying the possibility that the ferromagnetism of 
A$_{1-x}$La$_{x}$B$_{6}$ reported so far is neither intrinsic.
\end{abstract}


]
\narrowtext

\section{Introduction}
\label{sec:introduction}
Recently, ferromagnetism with high $T_{\rm C}$ ($\sim 600$ K) 
has been reported in alkaline-earth hexaborides doped with La, 
A$_{1-x}$La$_{x}$B$_{6}$ (A=Ca, Sr, and Ba). \cite{Young99} 
According to the study, the magnitude of ferromagnetic moment 
varies with La concentration, $x$, and is maximum at $x=0.005$, 
though very tiny ($< 10^{-3} \mu_{\rm B}$/unit cell). Since this 
series of compounds have no magnetic elements, the appearance of 
ferromagnetism is quite surprising and has stimulated a number of 
studies on the mechanism of ferromagnetism. The parent compound, 
alkaline-earth hexaboride, AB$_{6}$, contains the CsCl 
arrangement of divalent alkaline-earth ions and B$_{6}$ clusters, 
and early theoretical work on the cluster calculation of 
B$_{6}$ indicates that a B$_{6}$ cluster with a transfer of two 
electrons (from the divalent alkaline-earth ion) takes a closed-shell 
electronic structure. \cite{Higgins54} More detailed band 
calculations \cite{Hasegawa79,Massidda97,Rodriguez00} indicate 
that there is a band overlap at the X points of the Brillouin zone 
between the valence band formed by the B 2$p$ state and the 
conduction band strongly hybridized with the alkaline-earth $d$, 
and thus AB$_{6}$ is a semimetal. Many theoretical models 
\cite{Zhitomirsky99,Balents00,Barzykin00} for the ferromagnetism 
put their basis on an ``excitonic'' state of AB$_{6}$, where the 
electrons and the holes in a semimetal form triplet excitons as 
binding states. Extra electrons introduced by La substitution into 
such a state favors parallel spin configuration to gain paring energy 
of excitons, and yields a ferromagnetic state as a result.

One of the important aspects of these theories is that the magnetic 
properties of A$_{1-x}$La$_{x}$B$_{6}$ are dominated by carrier 
doping. Such a doping-dependent ferromagnetism is analogous to 
the ferromagnetism of perovskite manganites, and experimental 
studies of charge dynamics are indispensable to clarify such 
ferromagnetism caused by carrier doping. However, there has been 
little systematic investigation of the electronic structure of 
A$_{1-x}$La$_{x}$B$_{6}$ as a function of La concentration. It 
should be stressed here that to know how electronic states evolve 
with La doping is the first step to experimentally understand 
possible relationship between ferromagnetism and charge dynamics 
in A$_{1-x}$La$_{x}$B$_{6}$. 

Another important issue from the experimental viewpoint is how to 
characterize samples properly. Recent studies on 
A$_{1-x}$La$_{x}$B$_{6}$ suggests strong sample dependence as 
well as spatial inhomogeneity of 
ferromagnetism.\cite{Kunii99,Terashima00,Vonlanthen00,Urbano02} 
Here, we have to be careful about the fact that the magnitude of 
the ferromagnetic moment is so tiny and can be easily affected by a 
small amount of impurities. It is very important, therefore, to 
characterize samples properly in terms of carrier concentration, 
possible spatial inhomogeneity, and impurity.

In the present study, we carried out optical reflectivity 
measurement as well as Hall measurement of 
A$_{1-x}$La$_{x}$B$_{6}$ with systematically changing $x$. 
These measurements are very powerful technique to obtain basic 
parameters for charge dynamics, for example, the effective mass 
and the concentration of carriers. Furthermore, the inhomogeneity 
of the sample can easily be checked by utilizing microscopy 
technique of reflectivity measurement. The aim of the present study 
is to investigate the evolution of electronic structures as well as 
the variation of magnetic properties with La doping, and to clarify 
whether the charge dynamics is really related to the 
``ferromagnetism'' of A$_{1-x}$La$_{x}$B$_{6}$. 

\section{Experiment}
\label{sec:experiment} 
Single crystals of A$_{1-x}$La$_{x}$B$_{6}$ (A=Ca and Sr, $0 
\le x \le 0.02$) were grown by an Al flux method. 
CaCO$_{3}$ (4N), SrCO$_{3}$(4N), LaB$_{6}$ (3N), boron (5N), 
and Al (4N) were used as starting materials and flux. CaB$_{6}$ or 
SrB$_{6}$ was made by borothermal reduction, and was put into an 
alumina crucible together with LaB$_{6}$ and Al. The materials 
were heated up to 1500 $^\circ$C and slowly cooled down under Ar 
atmosphere. \cite{Synthesis} Plate-like samples with (100) surface 
with a typical dimension of 1 mm $\times$ 1 mm $\times$ 0.1 mm 
were obtained. The detail of the sample characterization is 
discussed in Sec.\ \ref{sec:magnetization}. Hall measurement was 
performed by applying magnetic field between -5 and 5 T. The 
electrode was attached by directly melting and bonding Au wire 
onto the sample surface. Reflectivity was measured between 0.07 
and 0.6 eV using a Fourier-transform infrared spectrometer 
equipped with a microscope. We checked the dependence of the 
spectra on surface treatment, and the result is discussed in details 
in Sec.\ \ref{sec:reflectivity}. Magnetization was measured by a 
superconducting quantum interference device (SQUID) 
magnetometer. Since the volume of each single crystal was too 
small, more than 10 pieces were combined for the magnetization 
measurement. In each measurement, a background signal was 
measured separately and subtracted from a total signal.

\section{Hall measurement}
\label{sec:Hall}
\begin{figure}
 \centerline{
  \epsfxsize=6.5cm 
  \epsfbox{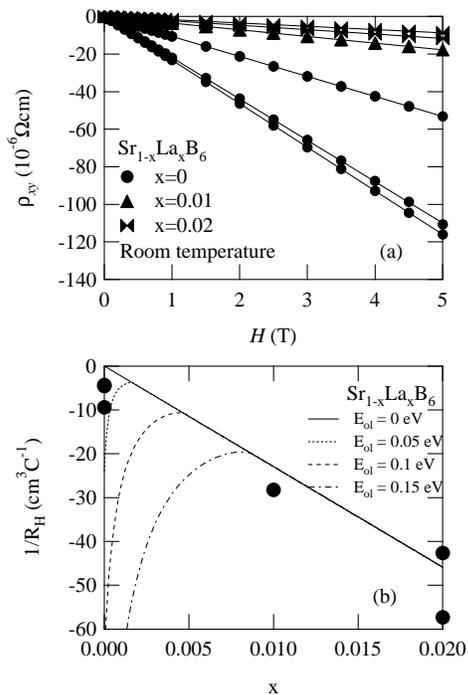}
 \vspace*{3mm}
 }
\caption{
(a) Hall resistivity vs magnetic field for 
Sr$_{1-x}$La$_{x}$B$_{6}$ at 300 K. The solid lines are 
least-square fits of the results to $\rho_{\rm xy}=R_{\rm H} H$. 
(b) Inverse Hall coefficients ($1/R_{\rm H}$) as a function of La 
concentration $x$ for Sr$_{1-x}$La$_{x}$B$_{6}$. The solid line 
shows the relation, $1/R_{\rm H}=xe$, and the dotted line, the 
dashed line, and the dot-dashed line show the calculated values of 
$1/R_{\rm H}$ with the band overlap ($E_{\rm ol}$) of 0.05 eV, 
0.1 eV, and 0.15 eV, respectively.
\label{fig:Hall}
}
\end{figure}

Hall resistivity $\rho_{\rm xy}$ vs. magnetic field $H$ for 
Sr$_{1-x}$La$_{x}$B$_{6}$ at room temperature is shown in 
Fig.\ \ref{fig:Hall}. 

In usual ferromagnetic metals, nonlinear 
behaviors of $\rho_{\rm xy} (H)$ are often observed (the so-called 
anomalous Hall effect), particularly around $T_{\rm C}$. This 
effect comes from the anomalous term of Hall resistivity 
proportional to magnetization, $R_{\rm s}M$. \cite{Rhyne68} 
However, $\rho_{\rm xy} (H)$ of 
Sr$_{1-x}$La$_{x}$B$_{6}$ shows a linear dependence with no 
sign of the anomalous Hall effect for any composition, as shown in 
Fig.\ \ref{fig:Hall}(a). 

When the anomalous term does not exist, Hall resistivity is given 
only by an ordinary term proportional to magnetic field, $R_{\rm 
H}H$. Figure \ \ref{fig:Hall}(b) plots the inverse Hall coefficient 
$1/R_{\rm H}$ as a function of $x$. The negative values of 
$R_{\rm H}$ mean that the majority carriers are electrons. The 
solid line gives the relation $1/R_{\rm H}=-xe$. The agreement 
between the experimental results and the simple relation $1/R_{\rm 
H}=-xe$ indicates that La doping introduces the same number of 
electrons into the conduction band. If the effect of band overlap in 
a semimetal is taken into account, the relation between Hall 
coefficient and $x$ becomes more complicated. This issue will be 
discussed in the next section.

\section{Reflectivity measurement}
\label{sec:reflectivity}
Hall coefficients are dominated only by the number of carriers 
$n$ but do not reflect their effective mass $m^{*}$. On the other 
hand, a reflectivity spectrum can reveal the value of 
$n/m^{*}$ through its plasma frequency, $\omega_{\rm p}=\sqrt{4 
\pi n e^{2}/\epsilon_{\infty}m^{*}}$, where 
$\epsilon_{\infty}$ is the dielectric constant at higher than the 
plasma frequency. 

\begin{figure}
\centerline{
  \epsfxsize=6.5cm 
  \epsfbox{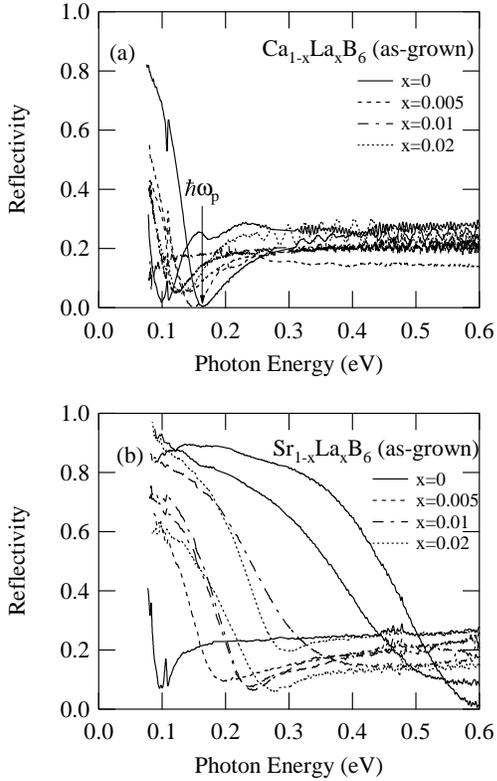}
 \vspace*{2mm}
 }
\caption{
Reflectivity spectra of (a) Ca$_{1-x}$La$_{x}$B$_{6}$ and (b) 
Sr$_{1-x}$La$_{x}$B$_{6}$ on as-grown surface. The arrow shows 
the position of the plasma edge of a typical spectrum.
\label{fig:ref_as_grown}
}
\end{figure}

Figure \ref{fig:ref_as_grown} shows the reflectivity spectra of 
Ca$_{1-x}$La$_{x}$B$_{6}$ and Sr$_{1-x}$La$_{x}$B$_{6}$ on 
as-grown surface as well as on slightly polished surface (by $<1 
\mu$m in depth). As can be seen, almost all the spectra have a clear 
plasma edge, as typically shown by an arrow, but the values of 
$\hbar \omega_{\rm p}$ with the same composition are fairly 
scattered. In fact, it is found that even pieces from the same 
crucible show different $\hbar \omega_{\rm p}$ values. As a result, 
a systematic variation of $\hbar \omega_{\rm p}$ with $x$, which 
is expected from the Hall-coefficient measurement, is barely 
observed. As an overall feature, the Sr series have larger values of 
$\hbar \omega_{\rm p}$ than the Ca series, similarly with the 
previous reports. \cite{Vonlanthen00,Ott97}

It should be noted here that reflectivity measurements detect only 
the sample surface with the penetration depth of light (the order of 
$\mu$m). Thus, if the sample surface with several $\mu$m in 
thickness has a different characteristic from the bulk one, the 
optical result can be inconsistent with the Hall measurement. 

\begin{figure}
\centerline{
  \epsfxsize=6.5cm 
  \epsfbox{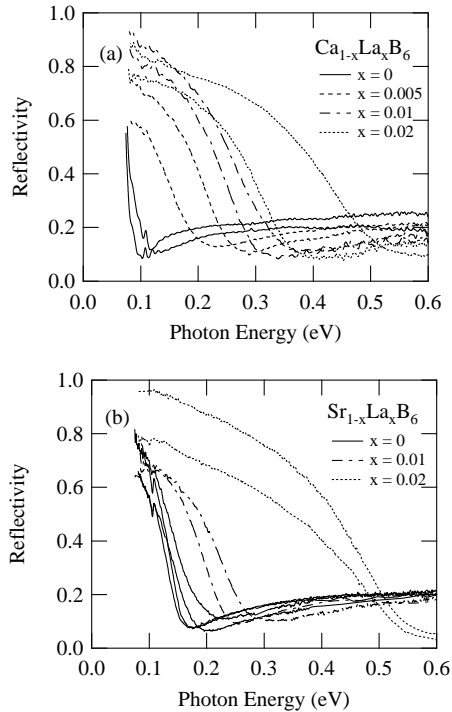}
 \vspace*{1mm}
 }
\caption{
Reflectivity spectra of (a) Ca$_{1-x}$La$_{x}$B$_{6}$ and (b) 
Sr$_{1-x}$La$_{x}$B$_{6}$ on the surface with filing ($\sim$ 10 
$\mu$m in depth) and polishing.
\label{fig:ref_filed}
}
\end{figure}

To check for this possibility, we have filed the sample surface by 
$\sim 10$ $\mu$m in depth, then polished it, and measured its 
reflectivity again. The reflectivity spectra after such surface 
treatment are shown in Fig.\ \ref{fig:ref_filed}. 
The scattering of the $\hbar \omega_{\rm p}$ values with the same
composition is drastically suppressed, and it is clearly observed
that $\hbar \omega_{\rm p}$ shifts to higher energy with increasing $x$.
We have also checked that there is almost no position dependence of 
the spectra along the sample surface.

\begin{figure}
\centerline{
  \epsfxsize=6.5cm 
  \epsfbox{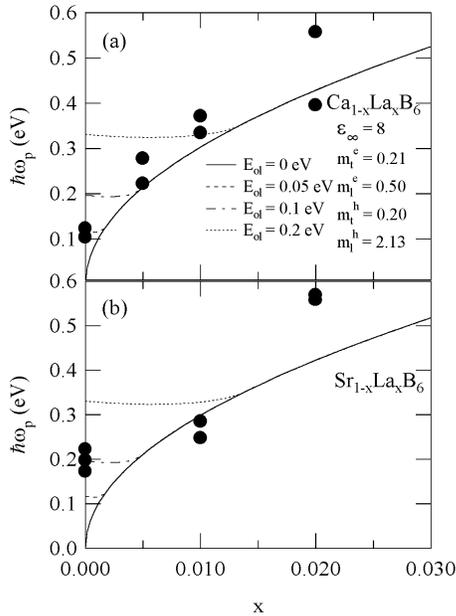}
 \vspace*{3mm}
 }
\caption{
Calculated values of $\hbar \omega _{\rm p}$ for several values of 
band overlap $E_{\rm ol}$ (lines), and the experimental values of 
$\hbar \omega_{\rm p}$ (closed circles) for (a) 
Ca$_{1-x}$La$_{x}$B$_{6}$ and (b) Sr$_{1-x}$La$_{x}$B$_{6}$.
\label{fig:x_omegap}
}
\end{figure}

To compare the experimental results with theoretical models, the 
$\omega_{\rm p}$ value has been calculated based on the 
parameters from a band calculation. \cite{Rodriguez00} First, holes 
in the valence band are ignored and only electrons in the conduction 
band are taken into account. The band structure of AB$_{6}$ has 
electron pockets at the triply degenerate X point [(100), (010), 
(001)], and the effective mass of the pocket is anisotropic between 
the longitudinal direction (parallel to the $\Gamma$X direction) 
and the transverse direction (perpendicular to the $\Gamma$X 
direction). According to a band calculation,  \cite{Rodriguez00} 
the longitudinal mass ($m_{\rm el}$) is $0.50m_{0}$ ($m_{0}$ is 
the free electron mass) and the transverse mass ($m_{\rm et}$) is 
$0.21m_{0}$. In this case, the plasma frequency $\omega_{\rm 
p}$ is given by the following equation,
\begin{equation}
\omega_{\rm p}
=
\sqrt{ 
\frac{4 \pi e^{2}}{\epsilon _{\infty }} 
\left(
\frac{n_{\rm e}/3}{m_{\rm el}} + \frac{2n_{\rm e}/3}{m_{\rm 
et}} 
\right)
},
\label{eq:plasma_frequency}
\end{equation}
where $n_{\rm e}$ is the number of electrons. 
$\epsilon_{\infty}$ is estimated to be 8 from a band calculation, 
which is consistent with the reflectivity value far above 
$\omega_{\rm p}$ in the experiment ($\sim 0.22$). The 
$x$ dependence of $\hbar\omega_{\rm p}$ calculated from 
Eq.\ \ref{eq:plasma_frequency} assuming $n_{\rm e}=x$ (solid 
lines) as well as the experimental values of $\hbar\omega_{\rm 
p}$ (closed circles) are plotted in Fig.\ \ref{fig:x_omegap}. The 
agreement between the experiment and the calculation is quite 
satisfactory (except for $x$=0 as discussed later), indicating that 
the effective mass of the conduction band by the band calculation 
describes the charge dynamics of these compounds correctly. 

One may notice an evident discrepancy at $x=0$ between the 
experiment (finite values of $\hbar \omega_{\rm p}$) and the 
calculation ($\hbar \omega_{\rm p}=0$.) To calculate the $\hbar 
\omega_{\rm p}$ value at $x \sim 0$, holes on the valence band, 
which exist even for $x=0$ in a semimetallic state, have to be taken 
into account. For the calculation, the effective mass of the valence 
band by the band calculation was used [the longitudinal mass 
($m_{\rm hl}$) is $2.13m_{0}$ and the transverse mass ($m_{\rm 
ht}$)  $0.20m_{0}$], but the band overlap ($E_{\rm ol}$) was 
taken as a free parameter. In this case, the plasma frequency is 
given by the sum of the contribution from electrons and holes as 
follows;
\begin{equation}
\omega_{\rm p}
=
\sqrt{ 
\frac{4 \pi e^{2}}{\epsilon _{\infty }} 
\left(
\frac{n_{\rm e}/3}{m_{\rm el}} 
+\frac{2n_{\rm e}/3}{m_{\rm et}}
+\frac{n_{\rm h}/3}{m_{\rm hl}} 
+\frac{2n_{\rm h}/3}{m_{\rm ht}}
\right) 
},
\label{eq:plasma_frequency2}
\end{equation}
where $n_{\rm h}$ is the number of holes, and $x=n_{\rm 
e}-n_{\rm h}$. The result of the calculation is shown in 
Fig.\ \ref{fig:x_omegap} for $E_{\rm ol} = 0.05$ eV (the dotted 
line) , 0.1 eV (the dashed line), and 0.2 eV (the dot-dashed line). 
From the comparison between the experiment and the calculation, 
the band overlap $E_{\rm ol}$ is estimated to be $\sim 0.05$ eV 
for CaB$_{6}$, and $\sim 0.1$ eV for SrB$_{6}$. 

A similar calculation taking account of both electrons and holes can 
be made for Hall coefficients. In this case, the total Hall 
coefficient is given by the subtraction of the hole contribution from 
the electron contribution. The result is shown in 
Fig.\ \ref{fig:Hall} (b), and from the comparison in Hall 
coefficients, $E_{\rm ol}$ is estimated to be $ < 0.05$ eV for 
SrB$_{6}$. The discrepancy of the $E_{\rm ol}$ values from 
reflectivity and Hall coefficients can be attributed to the deviation 
of the band structure from simple parabolic ones. 

\section{Magnetization measurement}
\label{sec:magnetization}
\begin{figure}
\centerline{
  \epsfxsize=6.5cm 
  \epsfbox{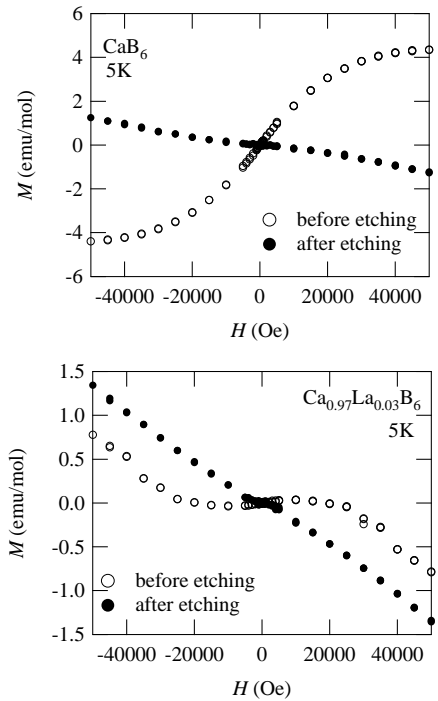}
 \vspace*{3mm}
 }
\caption{
Magnetization vs magnetic field at 5 K for CaB$_{6}$ (the upper 
panel) and Ca$_{0.97}$La$_{0.03}$B$_{6}$ (the lower panel) 
before (white circles) and after (closed circles) 
HNO$_{3}$ etching.
\label{fig:etch_mag}
}
\end{figure}

\begin{figure}
\centerline{
  \epsfxsize=6.5cm 
  \epsfbox{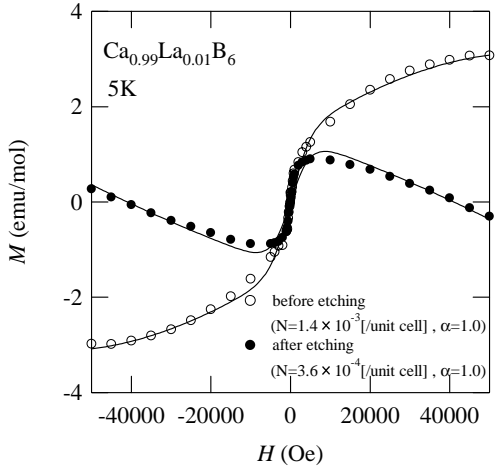}
 \vspace*{3mm}
 }
\caption{
Magnetization vs magnetic field for 
Ca$_{0.99}$La$_{0.01}$B$_{6}$ at 5 K before (white circles) and 
after (closed circles) HNO$_{3}$ etching. Solid lines are fitting 
results by eq.\ (3) (see text.) 
\label{fig:mag_fit}
}
\end{figure}

Since it has been clarified that the sample surface with 10 $\mu$m 
in thickness has a different electronic structure from the bulk one, 
the next question is how the magnetic properties of this part is 
different from the bulk magnetic properties. To answer this question, 
we first measured the magnetization of A$_{1-x}$La$_{x}$B$_{6}$, 
then etched the surface of the sample by HNO$_{3}$, and measured 
its magnetization again. Figure \ref{fig:etch_mag} shows the 
magnetization of Ca$_{1-x}$La$_{x}$B$_{6}$ as a function of 
magnetic field before and after etching the sample. A drastic change 
of the magnetization before and after etching is clearly observed. 
Roughly speaking, a positive component (a ferromagnetic or a 
paramagnetic component) decreases and a diamagnetic component 
survives with etching the sample surface. As most clearly seen in 
Fig.\ \ref{fig:mag_fit}, the positive component is likely composed 
of both a ferromagnetic, which saturates far below 10000 Oe, and a 
paramagnetic component, which gradually saturates up to 50000 Oe. 
To know which component changes most with etching, the 
magnetization curve has been fitted by the sum of three components, 
a paramagnetic, a ferromagnetic, and a diamagnetic component as 
follows:
\begin{eqnarray}
M = N g \mu _{\rm B} S B _{\rm S} (X)
&+&
\alpha M_{\rm ferro} (H)
+
\chi _{\rm dia} H,
\label{eq:magnetization}
\\
X 
&=&
g \mu _{\rm B} S H / k_{\rm B} T,
\nonumber
\end{eqnarray}
where $N$ is the number of paramagnetic moments, $g$ the g-factor 
of spin, $\mu _{\rm B}$ the Bohr magneton, $B_{\rm S}(X)$ the 
Brillouin function, $\alpha$ the amount of ferromagnetic moments, 
$M_{\rm ferro}(H)$ the magnetization curve for a ferromagnet, and 
$\chi _{\rm dia}$ the diamagnetic susceptibility of core electrons. 
$\chi _{\rm dia}$ was fixed to the value calculated from the 
diamagnetic susceptibility of Ca and six B ($\chi_{\rm dia}=-5.9 
\times 10^{-5}$ cm$^{3}$/mol). Figure \ref{fig:mag_fit} shows 
one of the fitting results. Here, $S=1/2$ for $B_{\rm S}(X)$ is 
adopted, which fits the data best. From this fitting, it is found that 
parameter $\alpha$, representing the amount of ferromagnetic 
moments, does not change by etching, but $N$, the number of 
paramagnetic moments, decreases to $\sim$ 40 \%. Similar results 
were obtained for other samples. Therefore, it can be concluded that 
the ferromagnetic moments are distributed over the sample 
uniformly, whereas the paramagnetic moments are localized at the 
sample surface.

\begin{figure}
\centerline{
  \epsfxsize=6.5cm 
  \epsfbox{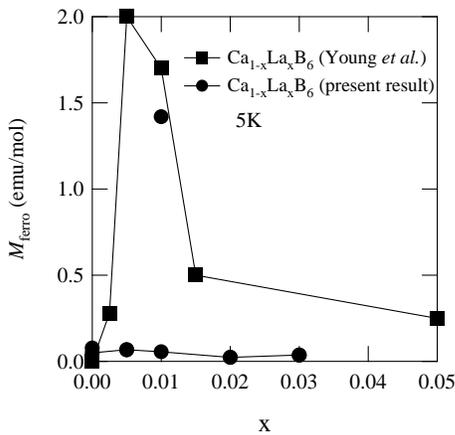}
 \vspace*{3mm}
 }
\caption{
Ferromagnetic moment vs La concentration $x$ for 
Ca$_{1-x}$La$_{x}$B$_{6}$. Closed circles correspond to the 
present result, whereas closed squares to the result in Ref.\ 1.
\label{fig:x_FMM}
}
\end{figure}

This result implies the importance of removing sample surface when 
one correctly estimates the bulk ferromagnetic moment of 
A$_{1-x}$La$_{x}$B$_{6}$. Therefore, we carefully removed the 
sample surface by HNO$_{3}$ etching, and measured the 
magnetization curve of A$_{1-x}$La$_{x}$B$_{6}$ with various 
values of $x$ and estimated the ferromagnetic moment. The size of 
the ferromagnetic moment as a function of $x$ in the present 
experiment is shown by closed circles in Fig.\ \ref{fig:x_FMM}, 
where the data from Ref.\ \onlinecite{Young99} are also plotted by 
closed squares. As can be seen, the ferromagnetic moment in the 
present experiment is substantially smaller than that of 
Ref.\ \onlinecite{Young99}, except for one sample ($x=0.01$), the 
same one shown in Fig.\ \ref{fig:mag_fit}. These results suggest 
that the ferromagnetic moment is not intrinsic but is caused by 
some impurities in the sample. 

\begin{figure}
\centerline{
  \epsfxsize=6.5cm 
  \epsfbox{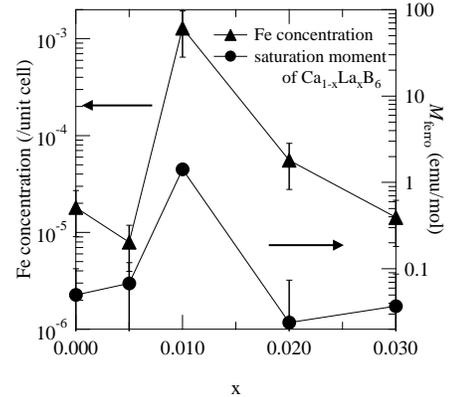}
 \vspace*{3mm}
 }
\caption{
Concentration of Fe impurity (closed triangles, left axis) and 
ferromagnetic moment (closed circles, right axis) as a function of 
$x$ for Ca$_{1-x}$La$_{x}$B$_{6}$.
\label{fig:ICP}
}
\end{figure}

To investigate what kind of and how much impurities exist in the 
sample, we took the following way. First, the magnetic impurities 
were searched qualitatively by X-ray fluorescence spectrometry. It 
was found from this technique that Fe is the main magnetic impurity 
in the sample. Then, we quantitatively determined the amount of Fe 
impurity by inductively coupled plasma atomic emission 
spectrometry (ICP-AES). Figure \ref{fig:ICP} shows the amount of 
Fe impurity (closed triangles), as well as the experimentally 
observed ferromagnetic moment of the same samples (closed 
circles). As can be seen, there is a rough correspondence between 
the amount of Fe impurity and the ferromagnetic moment. The 
x=0.01 sample with the largest ferromagnetic moment (as shown in 
Fig.\ \ref{fig:x_FMM}) turned out to be the one containing the 
largest amount of Fe impurity ($\sim 1000$ ppm, more than one 
order of magnitude larger than other samples). Furthermore, if we 
assume 1 $\mu _{B}$ moment per Fe, 7.3 emu/mol is expected in 
total, which exceeds the experimentally observed ferromagnetic 
moment, 1.4 emu/mol. On the other hand, other samples showing 
much smaller sizes of ferromagnetic moment (less than 0.1 
emu/mol) have much smaller amounts of Fe impurity (less than 50 
ppm), but they are also enough to produce the observed 
ferromagnetic moment. Therefore, we conclude that the 
ferromagnetic moment observed in the present experiment is caused 
by Fe impurity. \cite{Fe_impurity}

\section{Discussion}
\label{sec:discussion}
One of the conclusions from the reflectivity measurement is that 
A$_{1-x}$La$_{x}$B$_{6}$ can be regarded as simple doped 
semimetals, and no signature of an ``excitonic state'' has been 
observed in our reflectivity spectra for any $x$. Though we cannot 
completely exclude the possibility that such a feature exists out of 
our experimental range ($< 0.07$ eV), further evidence against the 
excitonic state in the doped samples can be derived in the following 
way: The calculated lines for the finite values of $E_{\rm ol}$ in 
Fig.\ \ref{fig:x_omegap} merge into the solid line (for $E_{\rm 
ol}=0$) at large $x$, where the Fermi level is located higher than 
the top of the valence band and the holes in the valence band are 
filled up. As can be seen in Fig.\ \ref{fig:x_omegap}, the 
$\omega_{\rm p}$ values in the experiment for $x \geq 0.005$ is 
in such a region, indicating that 0.5 \% La doping is enough to fill 
up the holes on the valence band. This indicates that the excitonic 
state, even if it exits for $x=0$, has already disappeared for 
$x=0.005$. 

Recent band calculation by the so-called GW approximation 
indicates that stoichiometric CaB$_{6}$ is not a semimetal but a 
narrow-gap band insulator,\cite{Tromp01} and an angle-resolved  
photoemission experiment indicates the existence of a band gap and 
the Fermi level located at the conduction band, resulting in only 
electron pockets. \cite{Denlinger01} These results are inconsistent 
with the interpretation of our experimental results based on a 
semimetal model. For this issue, we cannot exclude the possibility 
that the plasma edge for $x=0$ comes from the doped electrons into 
a conduction band of a band insulator. In other words, it is possible 
that the size of the band overlap is zero (solid lines in 
Fig.\ \ref{fig:x_omegap}), but the $x$ axis in 
Fig.\ \ref{fig:x_omegap} is shifted by $\sim 0.002$ for Ca series 
and by $\sim 0.005$ for Sr series because of offstoichiometry (most 
probably B defect). However, it should be pointed out that we 
measured a number of pieces of the same composition, and found 
that $\hbar\omega _{\rm p}$ is $\sim 0.1$ eV for all 
CaB$_{6}$ samples and $\sim 0.2$ eV for all SrB$_{6}$ samples. 
This can be easily explained by a semimetal model, as discussed in 
Sec.\ \ref{sec:reflectivity}. By the doped insulator model, on the 
other hand, we have to assume that the amount of defects barely 
depend on samples, which seems fairly unlikely.

It is also found from the reflectivity measurement that the sample 
surface with $\sim$ 10 $\mu$m in thickness has a different 
electronic structure from the bulk one. This phenomenon seems 
generic for single crystals of these compounds grown in Al flux, 
judging from the results of previous optical studies by other groups. 
\cite{Vonlanthen00,Ott97} Since the phonon peak in the optical 
spectra around 0.11 eV, which is assigned to an internal mode of the 
B$_{6}$ cluster, \cite{Degiorgi97} is the same in energy before 
and after surface treatment, the surface state should be close to the 
bulk A$_{1-x}$La$_{x}$B$_{6}$ in terms of crystal structure. 
Possible origins of the surface state are such as a slightly oxidized 
phase or an offstoiciometric phase precipitated at low temperatures 
during single-crystal growth. Whichever is the case, such an effect 
changes the Fermi level, or even changes the band structure, and 
thus, varies the plasma frequency. 

Let us move on to the magnetism of A$_{1-x}$La$_{x}$B$_{6}$. 
As discussed in Sec.\ \ref{sec:magnetization}, both a 
ferromagnetic and a paramagnetic component coexist in the 
magnetization. The existence of a paramagnetic component in 
addition to a ferromagnetic component has not been explicitly 
discussed so far, but has already been observed in various 
experiments, for example, high-field magnetizations in 
Ref.\ \onlinecite{Hall01}. It is found from the present experiment 
that the paramagnetic moments are confined to the sample surface. 
What is the origin of these paramagnetic moments at the surface? 
The amount of Fe impurity at the surface is estimated from ICP-AES, 
but the value is not large enough to explain the experimentally 
obtained size of the paramagnetic moment. It is reasonable to think 
that the electronic structure of the surface, which is different from 
the bulk one as shown in the reflectivity spectrum, is related to the 
appearance of the paramagnetic moment. One possibility is that the 
defect of Ca or B at the sample surface@yields a local magnetic 
moment, as suggested by a recent calculation.\cite{Monnier01} 
However, further studies are necessary to understand the origin of 
the paramagnetic moment at the sample surface.

Regarding the ferromagnetism of A$_{1-x}$La$_{x}$B$_{6}$, the 
conclusion of the present experiment is that there is no intrinsic 
bulk ferromagnetic moment in our samples, but there is 
ferromagnetic moments caused by Fe impurity. It should be stressed 
here again that our sample is well characterized in terms of carrier 
concentration, and it is unlikely that we missed the concentration 
range for an intrinsic ferromagnetic phase, if there is such a phase. 
Therefore, our best statement on this issue is that the 
ferromagnetism of A$_{1-x}$La$_{x}$B$_{6}$ is not dominated by 
carrier doping. A plausible explanation is that any 
``ferromagnetism'' of A$_{1-x}$La$_{x}$B$_{6}$ reported so far is 
caused by Fe impurity, as is the case for our samples.

\section{Summary}
\label{sec:summary}
We have investigated the charge dynamics of 
A$_{1-x}$La$_{x}$B$_{6}$ by reflectivity and Hall measurement. 
It is found that La doping introduces the same number of electrons 
into a semimetallic state, and its effective mass is consistent with a 
band calculation. No evidence of an excitonic state is observed, but 
the system should be regarded as a simple doped semimental. It is 
also found that the as-grown sample surface with $\sim$ 10 
$\mu$m in thickness has a different electronic structure from a 
bulk one. From magnetization measurements, it is found that this 
surface part contains a large number of paramagnetic moments. We 
have carefully measured the ferromagnetic moment of 
A$_{1-x}$La$_{x}$B$_{6}$ after removing the surface part by 
etching process, and have found that the ferromagnetic moments in 
our samples are substantially smaller than those observed so far. 
This result, together with a good correspondence between the size 
of the ferromagnetic moment and the amount of Fe impurity in the 
sample, indicates that the ``ferromagnetism'' of 
A$_{1-x}$La$_{x}$B$_{6}$ is not intrinsic, but most probably 
caused by Fe impurity.

\section{Acknowledgment}
\label{sec:acknowledgment}
We thank M.\ Nohara, Z.\ Hiroi, and S.\ Horii for their help with 
sample growth at the early stage of this study, and Z.\ Fisk, 
S.\ Fujiyama and M.\ Takigawa for helpful discussions. The present 
work was partly supported by a Grant-In-Aid for Scientific 
Research from Ministry of Education, Culture, Sports, Science and 
Technology, Japan.


\begin{references}
\bibitem{Synthesis} Another possible way of growing single 
crystals is to synthesize Ca$_{1-x}$La$_{x}$B$_{6}$ powder by 
borothermal reduction and to put it in Al flux. We believe that the 
difference should not affect the physical properties of the sample, 
as long as the carrier concentration is characterized correctly.
\bibitem{Fe_impurity} To investigate where these Fe impurities 
come from, we also determined the amounts of Fe impurity of the 
starting materials by ICP-AES. The amounts of Fe impurity are 6 
ppm for CaCO$_{3}$, 2 ppm for B, 1040 ppm for LaB$_{6}$, and 2 
ppm for Al. On the basis of these values, we can explain Fe 
impurity less than 50 ppm in the single crystals by assuming that 
the Fe impurity of starting materials was slightly condensed to the 
crystals in the process of flux growth. However, for the one with 
the largest ferromagnetic moment (1emu/mol), which has 1000 ppm 
Fe impurity, we guess that it is owing to an accidental 
contamination during synthesis.
\end{references}
\end{document}